\documentclass[twocolumn]{aastex62}
\graphicspath{{./}{figures/}}
\shorttitle{Correcting Turbulence in DESI}
\shortauthors{Schlafly et al.}
\usepackage{textcomp}
\usepackage{hyperref}
\usepackage{amsmath}
\usepackage{amsfonts}

\newcommand{\mum}{\ensuremath{\mathrm{\mu m}}}
\newcommand{\bs}[1]{\boldsymbol{#1}}

\begin{document}

\title{Correcting Turbulence-induced Errors in Fiber Positioning for the Dark Energy Spectroscopic Instrument}


\author[0000-0002-3569-7421]{E.~F.~Schlafly}
\affiliation{Space Telescope Science Institute, 3700 San Martin Drive, Baltimore, MD 21218, USA}

\author[0000-0001-9822-6793]{J.~Guy}
\affiliation{Lawrence Berkeley National Laboratory, 1 Cyclotron Road, Berkeley, CA 94720, USA}

\author[0000-0002-6550-2023]{K.~Honscheid}
\affiliation{Center for Cosmology and AstroParticle Physics, The Ohio State University, 191 West Woodruff Avenue, Columbus, OH 43210, USA}
\affiliation{Department of Physics, The Ohio State University, 191 West Woodruff Avenue, Columbus, OH 43210, USA}

\author[0000-0003-4207-7420]{S.~Kent}
\affiliation{Department of Astronomy and Astrophysics, University of Chicago, 5640 South Ellis Avenue, Chicago, IL 60637, USA}
\affiliation{Fermi National Accelerator Laboratory, PO Box 500, Batavia, IL 60510, USA}

\author[0000-0003-2644-135X]{S.~E.~Koposov}
\affiliation{Institute for Astronomy, University of Edinburgh, Royal Observatory, Blackford Hill, Edinburgh EH9 3HJ, UK}
\affiliation{Institute of Astronomy, University of Cambridge, Madingley Road, Cambridge CB3 0HA, UK}
\affiliation{Kavli Institute for Cosmology, University of Cambridge, Madingley Road, Cambridge CB3 0HA, UK}

\author{J.~Aguilar}
\affiliation{Lawrence Berkeley National Laboratory, 1 Cyclotron Road, Berkeley, CA 94720, USA}

\author[0000-0001-6098-7247]{S.~Ahlen}
\affiliation{Physics Dept., Boston University, 590 Commonwealth Avenue, Boston, MA 02215, USA}

\author[0000-0003-4162-6619]{S.~Bailey}
\affiliation{Lawrence Berkeley National Laboratory, 1 Cyclotron Road, Berkeley, CA 94720, USA}

\author{D.~Brooks}
\affiliation{Department of Physics \& Astronomy, University College London, Gower Street, London, WC1E 6BT, UK}

\author{T.~Claybaugh}
\affiliation{Lawrence Berkeley National Laboratory, 1 Cyclotron Road, Berkeley, CA 94720, USA}

\author{K.~Dawson}
\affiliation{Department of Physics and Astronomy, The University of Utah, 115 South 1400 East, Salt Lake City, UT 84112, USA}

\author{P.~Doel}
\affiliation{Department of Physics \& Astronomy, University College London, Gower Street, London, WC1E 6BT, UK}

\author[0000-0003-2371-3356]{K.~Fanning}
\affiliation{Kavli Institute for Particle Astrophysics and Cosmology, Stanford University, Menlo Park, CA 94305, USA}
\affiliation{SLAC National Accelerator Laboratory, Menlo Park, CA 94305, USA}

\author[0000-0003-2808-275X]{D. P. Finkbeiner}
\affiliation{Department of Physics, Harvard University, 17 Oxford St., Cambridge, MA 02138, USA}
\affiliation{Harvard-Smithsonian Center for Astrophysics, 60 Garden St., Cambridge, MA 02138, USA}

\author[0000-0002-3033-7312]{A.~Font-Ribera}
\affiliation{Department of Physics \& Astronomy, University College London, Gower Street, London, WC1E 6BT, UK}
\affiliation{Institut de F\'{i}sica d’Altes Energies (IFAE), The Barcelona Institute of Science and Technology, Campus UAB, 08193 Bellaterra Barcelona, Spain}

\author[0000-0002-2890-3725]{J.~E.~Forero-Romero}
\affiliation{Departamento de F\'isica, Universidad de los Andes, Cra. 1 No. 18A-10, Edificio Ip, CP 111711, Bogot\'a, Colombia}
\affiliation{Observatorio Astron\'omico, Universidad de los Andes, Cra. 1 No. 18A-10, Edificio H, CP 111711 Bogot\'a, Colombia}

\author[0000-0003-3142-233X]{S.~Gontcho A Gontcho}
\affiliation{Lawrence Berkeley National Laboratory, 1 Cyclotron Road, Berkeley, CA 94720, USA}

\author{G.~Gutierrez}
\affiliation{Fermi National Accelerator Laboratory, PO Box 500, Batavia, IL 60510, USA}

\author[0000-0002-8828-5463]{D.~Kirkby}
\affiliation{Department of Physics and Astronomy, University of California, Irvine, 92697, USA}

\author[0000-0003-3510-7134]{T.~Kisner}
\affiliation{Lawrence Berkeley National Laboratory, 1 Cyclotron Road, Berkeley, CA 94720, USA}

\author[0000-0001-6356-7424]{A.~Kremin}
\affiliation{Lawrence Berkeley National Laboratory, 1 Cyclotron Road, Berkeley, CA 94720, USA}

\author[0000-0003-2999-4873]{J.~Lasker}
\affiliation{Department of Physics, Southern Methodist University, 3215 Daniel Avenue, Dallas, TX 75275, USA}

\author[0000-0003-1838-8528]{M.~Landriau}
\affiliation{Lawrence Berkeley National Laboratory, 1 Cyclotron Road, Berkeley, CA 94720, USA}

\author[0000-0001-7178-8868]{L.~Le~Guillou}
\affiliation{Sorbonne Universit\'{e}, CNRS/IN2P3, Laboratoire de Physique Nucl\'{e}aire et de Hautes Energies (LPNHE), FR-75005 Paris, France}

\author[0000-0003-1887-1018]{M.~E.~Levi}
\affiliation{Lawrence Berkeley National Laboratory, 1 Cyclotron Road, Berkeley, CA 94720, USA}

\author[0000-0002-1769-1640]{A.~de la Macorra}
\affiliation{Instituto de F\'{\i}sica, Universidad Nacional Aut\'{o}noma de M\'{e}xico,  Cd. de M\'{e}xico  C.P. 04510,  M\'{e}xico}

\author[0000-0002-4279-4182]{P.~Martini}
\affiliation{Center for Cosmology and AstroParticle Physics, The Ohio State University, 191 West Woodruff Avenue, Columbus, OH 43210, USA}
\affiliation{Department of Astronomy, The Ohio State University, 4055 McPherson Laboratory, 140 W 18th Avenue, Columbus, OH 43210, USA}
\affiliation{The Ohio State University, Columbus, 43210 OH, USA}

\author[0000-0002-1125-7384]{A.~Meisner}
\affiliation{NSF NOIRLab, 950 N. Cherry Ave., Tucson, AZ 85719, USA}

\author{R.~Miquel}
\affiliation{Instituci\'{o} Catalana de Recerca i Estudis Avan\c{c}ats, Passeig de Llu\'{\i}s Companys, 23, 08010 Barcelona, Spain}
\affiliation{Institut de F\'{i}sica d’Altes Energies (IFAE), The Barcelona Institute of Science and Technology, Campus UAB, 08193 Bellaterra Barcelona, Spain}

\author[0000-0002-2733-4559]{J.~Moustakas}
\affiliation{Department of Physics and Astronomy, Siena College, 515 Loudon Road, Loudonville, NY 12211, USA}

\author[0000-0002-1544-8946]{G.~Niz}
\affiliation{Departamento de F\'{i}sica, Universidad de Guanajuato - DCI, C.P. 37150, Leon, Guanajuato, M\'{e}xico}
\affiliation{Instituto Avanzado de Cosmolog\'{\i}a A.~C., San Marcos 11 - Atenas 202. Magdalena Contreras, 10720. Ciudad de M\'{e}xico, M\'{e}xico}

\author[0000-0001-7145-8674]{F.~Prada}
\affiliation{Instituto de Astrof\'{i}sica de Andaluc\'{i}a (CSIC), Glorieta de la Astronom\'{i}a, s/n, E-18008 Granada, Spain}

\author{G.~Rossi}
\affiliation{Department of Physics and Astronomy, Sejong University, Seoul, 143-747, Korea}

\author[0000-0002-9646-8198]{E.~Sanchez}
\affiliation{CIEMAT, Avenida Complutense 40, E-28040 Madrid, Spain}

\author{M.~Schubnell}
\affiliation{Department of Physics, University of Michigan, Ann Arbor, MI 48109, USA}
\affiliation{University of Michigan, Ann Arbor, MI 48109, USA}

\author[0000-0003-3449-8583]{R.~Sharples}
\affiliation{Centre for Advanced Instrumentation, Department of Physics, Durham University, South Road, Durham DH1 3LE, UK}
\affiliation{Institute for Computational Cosmology, Department of Physics, Durham University, South Road, Durham DH1 3LE, UK}

\author{D.~Sprayberry}
\affiliation{NSF NOIRLab, 950 N. Cherry Ave., Tucson, AZ 85719, USA}

\author[0000-0003-1704-0781]{G.~Tarl\'{e}}
\affiliation{University of Michigan, Ann Arbor, MI 48109, USA}

\author{B.~A.~Weaver}
\affiliation{NSF NOIRLab, 950 N. Cherry Ave., Tucson, AZ 85719, USA}

\author[0000-0002-6684-3997]{H.~Zou}
\affiliation{National Astronomical Observatories, Chinese Academy of Sciences, A20 Datun Rd., Chaoyang District, Beijing, 100012, P.R. China}

\collaboration{(DESI)}

\begin{abstract}
Highly-multiplexed, robotic, fiber-fed spectroscopic surveys are observing tens of millions of stars and galaxies.  For many systems, accurate positioning relies on imaging the fibers in the focal plane and feeding that information back to the robotic positioners to correct their positions.  Inhomogeneities and turbulence in the air between the focal plane and the imaging camera can affect the measured positions of fibers, limiting the accuracy with which fibers can be placed on targets.  For the Dark Energy Spectroscopic Instrument, we dramatically reduced the effect of turbulence on measurements of positioner locations in the focal plane by taking advantage of stationary positioners and the correlation function of the turbulence.  We were able to reduce positioning errors from $7.3~\micron$ to $3.5~\micron$, speeding the survey by 1.6\% under typical conditions.
\end{abstract}


\section{Introduction} 
\label{sec:intro}

Massively-multiplexed, robotically-positioned, fiber-fed spectroscopic systems routinely survey the night sky, measuring the spectra of tens of millions of stars and galaxies.  Current and upcoming systems include LAMOST, SDSS-V, 4MOST, PFS, and MOONS \citep{Cui:2012, Kollmeier:2017, deJong:2019, Takada:2014, Cirasuolo:2014}, and we focus in this work on the Dark Energy Spectroscopic Instrument (DESI) \citep{Levi:2013}.  The robotic fiber positioners on these instruments can quickly position fibers at the locations of target astronomical sources.  Precise positioning of these fibers is important in order to maximize the amount of light received from targets.

Accurate positioning of fibers in the focal plane is challenging, however, due to tight tolerances: for example, the DESI design requires that positioners be placed within 10~\mum\ of targets.  Most systems contain guide cameras, which are used to bring the telescope to the correct position and the focal plane to the correct rotation.  The locations of stars in the guide cameras are then used to determine the target locations for fibers in the focal plane, using knowledge of the metrology of the focal plane and the mapping from the focal plane to the sky.  For DESI, our approach to positioning fibers is laid out in \citet{Kent:2023}.  The accuracy of the positioning is determined by a fiber dithering approach described in \citet{Schlafly:2024}, which we also use to improve the mapping between the focal plane and the sky.

Most current and upcoming multi-object spectrographs feature cameras that image the fibers in the focal plane \citep[e.g.]{Jurgenson:2020, Winkler:2022}.  These images are used to measure the locations of fibers, which are then used to improve their centering on target stars and galaxies.  These systems provide critical feedback about the quality of the positioning, but they introduce another set of optics into the problem, and require that the mapping between the focal plane and those cameras is well understood.  

\begin{figure}
    \centering
    \includegraphics[width=\columnwidth]{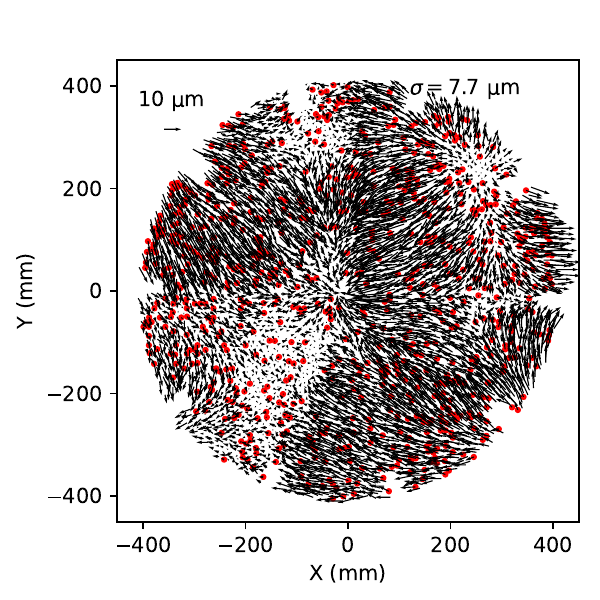}
    \caption{An example of turbulence-induced variations in the measured positions of fibers as seen by the Fiber View Camera (FVC) in DESI.  Arrows point from the average position over a long series of exposures to the measured position in a particular exposure.  The root-mean-square variation $\sigma = \sqrt{\langle x^2 + y^2 \rangle }$ of $7.7~\mum$ is given in the upper right, and an arrow of length $10~\mum$ is shown for comparison in the upper left. Red circles show the locations of stationary points which we will use to correct turbulence.}
    \label{fig:turbexample}
\end{figure}

When fibers are imaged in the focal plane for use in improving the positioning of fibers, turbulence from convective cells in the volume of air between the imaging camera and focal plane can distort the measured positions.  For DESI, inhomogeneities in the air imprint characteristic patterns on the observed positions of positioners, adding roughly 7~\mum\ of noise (0.1\arcsec) in the focal plane.  This is small compared to the DESI fiber diameter of 107~\mum\ (1.5\arcsec), but still reduces survey speed meaningfully.  Moreover, this value is large compared to the intrinsic precision of the motors (1.6~\mum, \citealt{Schubnell:2018}), so if we could correct the turbulence, we could position the fibers with much better accuracy.  Figure~\ref{fig:turbexample} shows an example of these turbulence-induced changes in the measured positions of DESI fibers.  The measurement errors are highly coherent from fiber to fiber, and therefore imprint correlated trends in the signal-to-noise ratio obtained on observed stars and galaxies if not corrected.

This work describes the approach taken by DESI to correct for the effects of turbulence on the measured positions of fibers.  The basic idea is to take advantage of the correlation function of the turbulence in combination with measurements of the locations of stationary points in the focal plane.  The stationary points provide measurements of the turbulence-induced errors in the positions, and a Gaussian process predicts the turbulent contribution to the measured positions at other locations.  Using this approach, we obtain a typical positioning error of 3.5~\mum, compared  with 7.3~\mum\ before correction.  This corresponds to 0.8\% more flux from a point source entering a fiber in typical 1\arcsec\ seeing, speeding the survey by 1.6\% in the background limited regime.

This work is organized as follows.  In \textsection\ref{sec:data}, we describe the DESI instrument and some of its data products, as well as the specific observations taken for this study.  In \textsection\ref{sec:method}, we describe the approach we take here to correct for the effect of turbulence on the measured positions of fibers.  In \textsection\ref{sec:results}, we discuss the results for DESI, the correlation function of turbulence observed, and the performance as a function of the available number of stationary points.  Finally, we conclude in \textsection\ref{sec:conclusion}.  The code and data used to produce the figures in this work are available at \url{https://zenodo.org/doi/10.5281/zenodo.12701420}.

\section{The DESI Instrument and Survey}
\label{sec:desi}

The DESI instrument resides at the prime focus of the Mayall telescope on Kitt Peak in Arizona.  The 4~m primary mirror focuses light through a wide-field optical corrector onto the focal plane \citep{Miller:2023}.  The focal plane is outfitted with 5000 fibers which can be robotically positioned in a 1.4\arcmin\ patrol radius to collect light from target stars and galaxies \citep{Silber:2023, Myers:2023}.  This light is then piped to ten spectrographs which cover 360~nm -- 980~nm \citep{desiinstrumentation:2022}.  The resulting spectra are extracted and calibrated to determine the spectral energy distributions of astronomical sources \citep{Guy:2023}.

The DESI collaboration is currently performing a survey of tens of millions of stars and galaxies \citep{Levi:2013, desi:2016a, Schlafly:2023}, selected from imaging from the DESI Legacy Imaging Survey \citep{Dey:2019}.  Primary target classes include Milky Way stars \citep{Cooper:2023}, bright galaxies \citep{Hahn:2023}, luminous red galaxies \citep{Zhou:2023}, emission line galaxies \citep{Raichoor:2023}, and quasars \citep{Chaussidon:2023}.  The first results from survey validation \citep{desisv} and DESI's Early Data Release \citep{desiedr} are now publicly available.  The survey recently reported measurements of the baryon-acoustic oscillation signal from the first year of DESI observations of galaxies \citep{desi:2024:3} and the Lyman-$\alpha$ forest \citep{desi:2024:4}, and the associated cosmological implications \citep{desi:2024:6}. 

When observing a new field, the DESI system performs several steps involving different instruments, which we describe below.  First, the telescope slews into place and, in parallel, the DESI positioners begin moving to their target locations in the focal plane.  Next the guide-focus array cameras (GFAs) identify bright stars and the telescope boresight is adjusted to precisely center and rotate the field.  After the initial positioner move is complete and the field has been acquired, the fibers are back-illuminated from the spectrographs and imaged by the Fiber-View Camera (FVC).  The FVC is located in the hole in the center of the Mayall primary mirror and looks through the corrector at the focal plane \citep{Baltay:2019}.  The current locations of fibers are measured using the FVC image, and used together with updated information about the mapping of the focal plane to the sky from the GFAs to determine how best to adjust the location of each fiber to center it on target sources.  The DESI positioners then execute a second move to bring the positioners closer to their target locations.  A final image of the back-illuminated fibers is taken to measure the locations of the positioners, and then spectroscopic observations begin.  For a more complete description of the positioning of DESI fibers, see \citet{Kent:2023}.

As described above, each time DESI positions fibers to observe a field, it moves the fibers twice.  The first move is termed the ``blind'' move because the positioners are simply commanded to go to a specific target location.  The second move is termed the ``correction'' move because it relies on feedback from the FVC to slightly improve the positioning of each fiber.

The FVC measurements of the fiber locations are the source of the data used in this work, so we briefly summarize some of \citet{Baltay:2019} to provide more detailed background information on the FVC and its optical path.  The FVC is located in the hole in the center of the Mayall primary mirror and looks through the corrector \citep{Miller:2023} at the focal plane.  The corrector includes six optical elements, including the atmospheric dispersion corrector (ADC).  The ADC's two optical elements move according to the airmass of the field being observed, to account for varying levels of dispersion.  The FVC features a 6k$\times$8k CCD with a plate scale of roughly 150~\mum\ in the DESI focal plane per FVC pixel.  The FVC aperture was chosen to give a FWHM of roughly 1.6 pixels, to allow for good centroiding.  Observed fiber locations are reported by fitting each fiber spot with an elliptical Gaussian, the center of which is the determined position.  FVC image processing includes bias and dark corrections, but no flat fielding is performed, as the flat field did not have enough small-scale structure to impact centroiding.  Outside a small region of the FVC that looks through a defect in the front surface of C3 \citep{Miller:2023}, the PSF does not vary substantially spatially.  The DESI positioning is substantially worse in that small region due to unmodeled small-scale distortions imparted by the defect \citep{Schlafly:2024}.  The PSF size is largely determined by diffraction and does not vary significantly with time.  The FVC measures centroids with an accuracy of 2~\mum, much smaller than the 7~\mum\ positioning error that was typical in DESI before we began correcting for the effect of turbulence.

\section{Data}
\label{sec:data}

This work focuses on using measurements of stationary points to measure and correct for the turbulence present in images from the FVC.  These stationary points include 120 fiducials \citep{desiinstrumentation:2022} exclusively used for calibration purposes, as well as several hundred non-functional positioners that are rarely used in operations.  For this work, we used a total of 793 stationary points, provided by the fiducials and 673 non-functional positioners; the locations of these points are shown in red in Figure~\ref{fig:turbexample}.  The non-functional positioners have a fairly random distribution in the focal plane.  Most of the non-functional positioners can move, but their movement is erratic and can lead to collisions with neighboring positioners, and so we have chosen to keep their locations fixed in DESI operations.  We expect to be able to return many of these positioners to operations in the future.  Fortunately, however, these stationary positioners provide good measurements of the sky background for DESI spectroscopy and can also be used to probe the turbulent positioning errors that are the focus of this work.

Every DESI observation results in a ``coordinates'' file\footnote{See the DESI data model at \url{https://desidatamodel.readthedocs.io/en/latest/DESI_SPECTRO_DATA/NIGHT/EXPID/coordinates-EXPID.html} for details.} which contains the measurements of the locations of fibers in each FVC exposure, both before and after turbulence correction.  These measurements are the focus of the analysis of this paper.

We use two sets of coordinates files in this work.  To get a broad sense for the accuracy of DESI positioning, we use all $\approx$10,000 DESI science observations since turbulence correction was enabled in December 2021.  To get a more detailed, higher-resolution look at turbulence, we also took a series of 400 FVC images of the back-illuminated focal plane, where all fibers were held fixed in place.  We use these observations to better measure the correlation function of the turbulence observed by DESI.  These images used the standard FVC exposure time (2~s) with the Mayall pointed directly at zenith.  This data is the basis for the analysis of \textsection\ref{subsec:correlation}.  By keeping all of the fibers fixed in place in this set of observations, we are able to use all fibers to measure turbulence, rather than only the stationary points.


\section{Method}
\label{sec:method}

We model the turbulence observed by the FVC using a Gaussian process, as we will describe in \textsection\ref{subsec:gp}.  The Gaussian process uses measurements of turbulence from a number of stationary points (\textsection\ref{subsec:stationary}) which are not moved during normal survey operations.  Thus, their apparent motion in FVC images can be largely attributed to turbulence, and the Gaussian process modeling conceptually averages and interpolates among them in an optimal way.  Some additional calibration of the stationary points is needed to make this procedure work well; we describe the calibration we perform in \textsection\ref{subsec:stationaryvariation}.

\subsection{Gaussian Process Modeling}
\label{subsec:gp}

Gaussian processes naturally describe the turbulence seen by DESI fibers.  We have measurements of the positions $\boldsymbol{x}_n$ of roughly 5000 fibers $n$ in two dimensions\footnote{Technically, we are counting here both the actual fibers and the fiducials.}.  These measurements derive from the centroids of the light of the back-illuminated fibers in images from the FVC.  The centroids are converted into locations in the focal plane using information about the current setting of the ADC, the measured locations of 120 fiducials, and a model for the optics of the DESI corrector lens.  See \citet{Kent:2023} for a more complete description of the process used to connect positioners with locations on the sky.  Using the nominal locations of stationary points (\textsection\ref{subsec:stationary}), we can compute the displacement $\boldsymbol{d}$ of the stationary points due to turbulence and noise.  The displacements $\boldsymbol{d}$ have a correlation function $C_{nm}$ which we model as the sum of a turbulent kernel $K$ and a diagonal noise term:
\begin{equation}
    C_{nm} = \langle \boldsymbol{d}_n \boldsymbol{d}_m \rangle = K(\boldsymbol{x}_n, \boldsymbol{x}_m) + \delta_{nm} \sigma^2 \, ,
\end{equation}
where $\delta_{nm}$ is the Kronecker delta and $n$ and $m$ index over fibers.  Given the correlation function $C$ and measured displacements $\boldsymbol{d}_n$, the predicted turbulence at any position $\boldsymbol{x}_s$ is given by
\begin{equation}
    \boldsymbol{d}_s = \sum_{nm} K(\boldsymbol{x}_s, \boldsymbol{x}_m) (C^{-1})_{nm} \boldsymbol{d}_n \, ,
\end{equation}
as described in \citet{Rasmussen:2006}.  Here $n$ and $m$ both loop over the indices of the measured fibers.  We follow Algorithm 2.1 of \citet{Rasmussen:2006} to solve for the maximum likelihood turbulence measurements $\boldsymbol{d}_s$.

This procedure relies on knowledge of the turbulence kernel $K(\boldsymbol{x}_n, \boldsymbol{x}_m)$, which describes how the turbulent offset at one location is correlated with the turbulent offset at another location.  We discuss this quantity more in \textsection\ref{subsec:correlation}.  For DESI, we took the approach of treating the turbulent offsets in the $x$ and $y$ direction as two separate, independent Gaussian fields, and chose a fixed squared exponential kernel for $K$:
\begin{equation}
    K(\boldsymbol{x}_n, \boldsymbol{x}_m) = A^2 \exp\left({-\frac{(\boldsymbol{x}_n - \boldsymbol{x}_m)^2}{2l^2}}\right) \, , 
\end{equation}
In principle we could have chosen to fit the parameters $A$, $l$, and $\sigma$ for each DESI exposure, but we opted to fix the terms of $C$ in order to speed the computation and simplify the algorithm.  We adopted $A = 5~\mum$, $l = 50~\mathrm{mm}$, and $\sigma = 5~\mum$ as a rough approximation to the observed correlation function and uncertainties.  However, in retrospect we overestimated the noise $\sigma$ and underestimated the amplitude $A$; see \textsection\ref{subsec:correlation} for more discussion.

Though we chose to model the turbulent offsets in the $x$ and $y$ components of $\boldsymbol{d}$ as independent Gaussian processes, other approaches better reflect the physics of the problem.  Because the absolute deflections are small, the location of a source in an image is given by the gradient of the projection of the density fluctuations in the air column onto a surface.  Accordingly, a different approach would be to model the surface as a scalar field with correlated variations across the focal plane.  The observed centroids would then be computed as the gradient of that scalar.  In that case, we have instead
\begin{equation}
\label{eq:gradwavefront}
K((\boldsymbol{x}_n)_i, (\boldsymbol{x}_m)_j) = \partial_i \partial_j K^\prime(\boldsymbol{x}_m, \boldsymbol{x}_n)
\end{equation}
where $i$ and $j$ index over the $x$ and $y$ components of $\boldsymbol{x}_n$, $K$ is the kernel for the observed displacements, and $K^\prime$ is the kernel for the scalar field whose gradient gives the observed displacements.  We do not use this approach for DESI operations, but we find in \textsection\ref{sec:results} that it performs slightly better than the independent approach.  We again use a squared exponential kernel for $K^\prime$ in this case, with $\sigma=2~\mum$, $A=1~\mathrm{mm}^2$, and $l=100~\mathrm{mm}$ for $K^\prime$.  The values for $A$ and $l$ are are different from those for the independent case because they describe the correlation of the scalar field rather than its derivative.  The value of $\sigma$ is different because it was developed later after better estimates of the uncertainties had been developed.

Occasionally DESI positioners may collide with their neighbors.  When a stationary positioner is bumped, its location can be significantly changed.  We do not want this movement to be interpreted as turbulence.  To avoid this, we added outlier rejection to the turbulence modeling.  We do this by first computing a simple median and clipped standard deviation of the displacements $d$.  We discard any points more than 5 standard deviations from the median.  We do an initial Gaussian process analysis, and further remove any points with measured locations more than 10~\mum\ from the predictions from the turbulence modeling.  Finally, we repeat the Gaussian process modeling with the remaining points.  This is {\it ad hoc} but has delivered good performance even during rare occurrences when we inadvertently bumped several nearby stationary positioners.

\subsection{Stationary Points}
\label{subsec:stationary}

The key measurements needed for the Gaussian process modeling of \textsection\ref{subsec:gp} are the displacements of the stationary points due to turbulence.  Because the stationary points do not move, in principle these measurements can be made by averaging together their observed locations over a long series of ordinary observations.  In practice, we have instead dedicated a small number of nights to measuring the locations of the stationary points.  These measurements can be made when bad weather prevents scientific observations.  We observe small changes to the ``stationary'' points ($<2~\mum$) from month to month, but these offsets are not large enough to meaningfully affect the survey speed.  Additionally, although infrequently, a handful of positioners are bumped and have significantly altered positions.  These positions are corrected as part of the routine updates to the stationary points, and are usually large enough that the outlier clipping of \textsection\ref{sec:method} identifies and removes them.

For DESI, the stationary points come in two varieties.  Most of the stationary points (673 in this work) come from back-illuminated fibers on non-functional positioners that are fixed in location and measured by the FVC.  The remaining stationary points (120) come from fiducials, which contain an LED which shines through four pinholes with a characteristic pattern to the FVC.  DESI reports an average position for each fiducial in each image, averaging over the four pinholes.  The fiducials in principle provide somewhat better positional measurements than the non-functional fibers, since they are mechanically more stable and produce four points in the FVC images, but we do not find significant differences in accuracy between the fiducials and non-functional positioners.

\subsection{Variation with Telescope Position}
\label{subsec:stationaryvariation}

Though we never intentionally move the stationary points, we find that their apparent positions move systematically depending on the direction the telescope is pointing on the sky.  The root-mean-square motion is 4.5~\mum\ and is clearly detected in the motions of the stationary points following long slews.  These motions are spatially coherent across the focal plane.  Moreover, both non-functional positioners and fiducials show the same coherent patterns, and so we expect that these apparent motions are an effect in the optics rather than mechanical sag of individual positioners or petals.

To measure the apparent motion of stationary points, we took special FVC observations when the dome was closed due to bad weather.  We took a series of FVC images across a broad range of telescope hour angles and declinations, using ten times the normal FVC exposure time in order to reduce the effect of turbulence.  We then fit the apparent focal plane coordinates of each fiber with a cubic function of hour angle and declination.  The resulting fits have root-mean-square residuals around 2.7~\mum.  These fits were removed from the observed positions of the stationary points prior to fitting for turbulence, for all FVC location measurements.

\section{Results and Discussion}
\label{sec:results}

We began correcting positioning for the effects of turbulence in December 2021.  Since then, we have corrected more than 10,000 positioning sequences.  We discuss the results of this correction in \textsection\ref{subsec:posstats}, and then investigate the turbulence correlation function in \textsection\ref{subsec:correlation}.  In \textsection\ref{subsec:howmany}, we look at how the accuracy of the turbulence correction depends on the number of stationary points.

\subsection{Turbulence \& its Correction for DESI}
\label{subsec:posstats}

\begin{figure*}
    \centering
    \includegraphics[width=\textwidth]{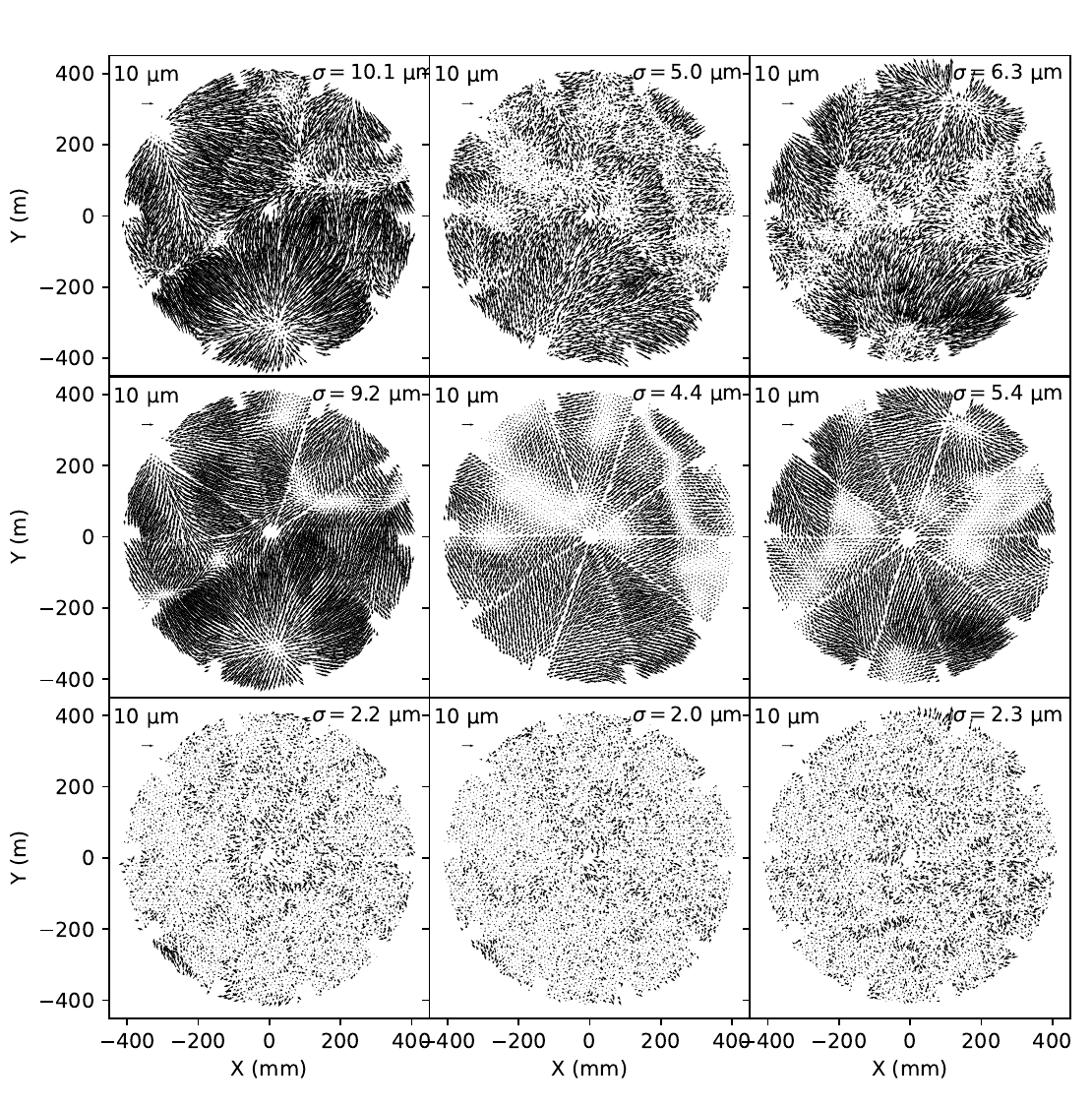}
    \caption{Three examples of turbulence.  Each column corresponds to a different exposure and therefore realization of the turbulence-induced position offsets.  The top row shows the measurements; the middle row shows the turbulence derived using the DESI turbulence algorithm; and the bottom row shows the residuals.  The turbulence modeling explains most of the variation and is able to reduce the residuals to roughly $2~\mum$.  The root-mean-square displacement is given in the upper right of each panel, and a 10~\mum\ scale bar is shown in the upper left.}
    \label{fig:turbulencegallery}
\end{figure*}

The effect of turbulence on positioning is immediately apparent in FVC measurements.  Figure~\ref{fig:turbulencegallery} shows three examples of images from a sequence of 400 FVC images taken with all positioners fixed in place.  The top row shows how the positions appear to move from their mean positions over the sequence due to turbulence.  The middle row shows the inferred turbulence field using the method of \textsection\ref{sec:method}.  Finally, the bottom row shows the residuals after removing the turbulence field.  While the observed displacements can be 10~\mum\ in size, after correction they are close to 2~\mum; most of the turbulence is fit and removed.  The residuals do show measurable remaining correlations at small spatial scales that we are not correcting, though 2~\mum\ residuals are too small to impact survey speed meaningfully.

\begin{figure}
    \centering
    \includegraphics[width=\columnwidth]{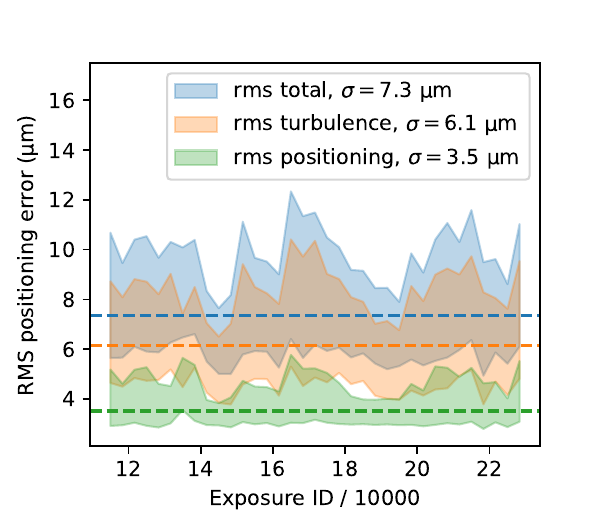}
    \caption{Fiber positioning performance for DESI.  The blue, orange, and green regions show the range between the 16th and 84th percentiles for DESI positioning accuracy over roughly 10,000 DESI positioning sequences. Darks, biases, and other exposures that do not involve a positioning loop are not used.  Absent any correction, the blue region shows that we would obtain typical precision of $7.3~\mum$.  We however fit for the turbulence in each image (orange), which contributes most of the variance.  The final positioning errors after correction for turbulence (green) have typical values of $3.5~\mum$.  Dashed horizontal lines show the median RMS positioning error over all of the exposures.}
    \label{fig:positioningstats}
\end{figure}

This level of turbulence is fairly typical of DESI observations.  Figure~\ref{fig:positioningstats} shows the positioning performance of DESI over 10,000 positioning sequences since DESI began computing turbulence corrections in December 2021.  The shaded regions show the 16th through 84th percentiles of the rms displacements in total (blue), due to turbulence (orange), and after correction for turbulence (green).  The dashed lines show the overall median of each quantity over all of the positioning sequences.

Figure~\ref{fig:positioningstats} shows that turbulence is the dominant contribution to the positioning error in each exposure, and that we remove most of it.  The total positioning error in the absence of turbulence corrections has a typical value of 7.3~\mum, of which 6.1~\mum\ is due to turbulence.  After correction for turbulence, the remaining residuals are 3.5~\mum.  This remaining error is due to a combination of imperfect motion of the positioners, noise in the measurements of the FVC, uncorrected turbulence, and unmodeled motions of the positioners when the telescope is moved (\textsection\ref{subsec:stationary}).  When the positioners are not moved, we can obtain residuals of $\sim 2$~\mum, which includes noise in the FVC measurements and uncorrected turbulence.  The DESI motors have a precision of $\approx 1.6$~\mum\ \citep{Schubnell:2018}.  Our expectation then is that the total positioning RMS in operations should be the quadrature sum of the zenith performance ($2~\mum$), the motor performance ($1.6~\mum$), and the performance at different hour angles and declinations (2.7~\mum).  The result of 3.7~\mum\ is close to the observed 3.5~\mum\ RMS performance. 

\subsection{The Turbulence Correlation Function}
\label{subsec:correlation}

The optimal correction for turbulence depends on the correlation function of the displacements it induces in measurements.  The correlation function of turbulence has been studied extensively.  For homogeneous, isotropic turbulent displacement fields, the correlation function of the displacement vector $\bs{d}$ as a function of the separation vector $\bs{r}$ can be fully characterized by the longitudinal ($\Psi_\parallel$) and transverse ($\Psi_\perp$) correlation functions \citep{Monin:1975}.  Note that ``parallel'' and ``perpendicular'' in this context refer to the direction of the turbulent displacements seen by two fibers relative to the separation between those fibers.

We measure $\Psi_\parallel$ and $\Psi_\perp$ for DESI by using a sequence of 400 DESI FVC observations for which the fibers were not moved, so that all of their apparent motion can be attributed to turbulence and uncertainty in the centroids.  We then have roughly 5000 measured displacements $\boldsymbol{d}_n$ for each of 400 exposures, where the separations between fibers is essentially constant.  We compute the correlation functions $\Psi_\parallel$ and $\Psi_\perp$ by considering all pairs of displacements $\boldsymbol{d}_n$ and $\boldsymbol{d}_m$, and computing their separation $\boldsymbol{r}_{nm}$.  The displacements have components parallel to $\boldsymbol{r}_{rm}$ ($d_\parallel$) and perpendicular to it ($d_\perp$).  We compute those components and then average their product together for all pairs with separations falling into a particular bin.  We end up with $\Psi_{\parallel k}$ and $\Psi_{\perp k}$, a binned approximation to the correlation functions $\Psi(r)$ we want to measure.  Mathematically, this computation is described as follows:
\begin{align}
    \Psi_{\parallel k} &= \frac{1}{N_k} \sum_{n>m} {d}_{\parallel,n}(\bs{r}_{nm}) {d}_{\parallel,m}(\bs{r}_{nm})I_k(\bs{r}_{nm})  \\
    \Psi_{\perp k} &= \frac{1}{N_k} \sum_{n>m} {d}_{\perp,n}(\bs{r}_{nm}) {d}_{\perp,m}(\bs{r}_{nm})I_k(\bs{r}_{nm}) \\
    \bs{r}_{nm} &= \bs{x}_n - \bs{x}_m \\
    {d}_{\parallel}(\bs{r}) &= \frac{\bs{d} \cdot \bs{r}}{|\bs{r}|} \\
    {d}_{\perp}(\bs{r}) &= \frac{|\bs{d} \times \bs{r}|}{|\bs{r}|} \\
    N_k &= \sum_{n > m} I_k(\boldsymbol{r}_{nm}) \\
    I_k(\boldsymbol{r}) &= \begin{cases}
    1 \quad \text{if } |\boldsymbol{r}| \text{ falls in bin } k \\
    0 \quad \mathrm{otherwise}
    \end{cases} \quad .
\end{align}
The bins are labeled by $k$ and the indicator function $I_k$ defines each bin.  We choose to make bins 1~mm in size, such that $I_k$ corresponds to separations from $k-1$~mm to $k$~mm.  In simple terms, we examine all pairs of stationary points with separations of a given distance.  We then calculate the average of the product of the parallel and perpendicular components of the displacements for each pair.  This process yields the parallel and perpendicular correlation functions.

\begin{figure}
    \centering
    \includegraphics[width=\columnwidth]{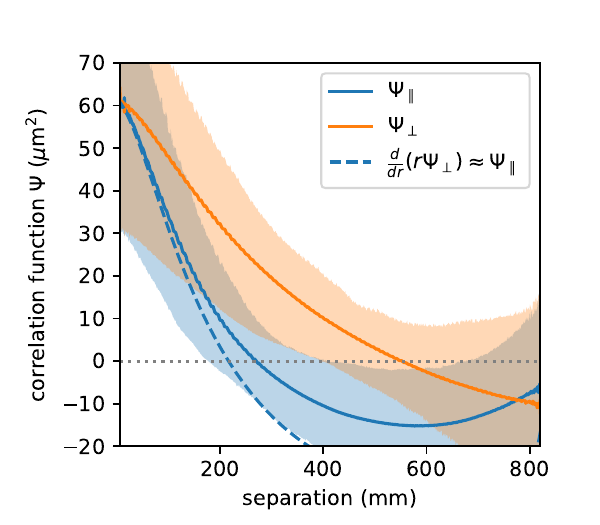}
    \caption{The correlation between positioning offsets along the direction of separation ($\Psi_\parallel$, longitudinal) and perpendicular to it ($\Psi_\perp$, transverse).  The quantity $\frac{d}{dr}(r\Psi_\perp)$ is shown by a dashed line, and should match $\Psi_\parallel$ if the turbulence is sourced by a scalar field.  There is good agreement at small separations though the agreement becomes worse at large separations; see text for discussion.  The shaded regions corresponds to the 16th to 84th percentile of the correlation function at each separation over the 400 images considered, while the solid lines give the average.
    }
    \label{fig:correlation}
\end{figure}

Figure~\ref{fig:correlation} shows the correlation functions in the parallel and perpendicular directions obtained from the average of 400 FVC observations.  The amplitude of the correlation function at zero separation is $(7.7~\micron)^2$, consistent with expectations given the total standard deviation of the measurements in the sample.  The correlation falls rather slowly, becoming negative near 300~mm for the longitudinal and 600~mm for the transverse correlation.  The parallel correlation falls roughly twice as quickly as the perpendicular correlation, and the parallel correlation halves near 100~mm.  This is a rather large scale; the radius of the DESI focal plane is only 420~mm, so not many turbulence measurements are needed to correct for most of the turbulence.

It may seem peculiar for the correlation function to become negative.  This is likely because the positions we measure have already had some filtering applied before our analysis.  DESI's PlateMaker package \citep{Kent:2023} fits out an overall scale, translation, and rotation in order to account for any shifts in the focus, position, or rotation of the FVC or focal plane.  These terms also naturally absorb any very large scale turbulence signal.

When the vector field can be expressed as the gradient of a scalar function, then it can be shown that $\frac{d}{dr}(r\Psi_\perp) = \Psi_\parallel$ \citep[e.g.][]{Gorski:1988}.  For DESI, we expect that the observed displacements are the gradient of the surface of projected air density fluctuations, so we expect good agreement here.  Figure~\ref{fig:correlation} shows the prediction for $\Psi_\parallel$ from this relation as the dashed line.  It matches the observed measurement closely at small separations but diverges at large separations, likely because of the extra filtering PlateMaker applies on large scales.  As a further test, we took several images with large turbulence and decomposed the displacement pattern into E and B modes using code that was developed for modeling distortions in the corrector optics \citep{Kent:2023}.  As expected, virtually all the effect (98\% to 99\%) is in E modes, and the small amount in B modes is likely due to noise.

\subsection{How many turbulence probes are needed?}
\label{subsec:howmany}

The accuracy of the turbulence correction depends on the number of stationary points that are available to probe the turbulence field.  How many are needed to obtain a given level of accuracy?  Judging by the correlation function (\textsection\ref{subsec:correlation}) we expect most of the turbulence to be corrected even when including only a small number of points, on the order of tens.  This follows because the correlation scale is roughly 100~mm and the focal plane is roughly 400~mm, and $(\frac{400}{100})^2 = 16$.  The total impact of turbulence on DESI's survey speed is only about $2\%$, and so a hypothetical correction which eliminated all turbulence would be worth about 2\% of all of DESI's fibers---100 fibers.  So were we able to pick a number of fibers to deactivate and use as turbulence probes, we would likely choose a number on the order of 20, and anything in excess of 100 would be clearly wasteful.

\begin{figure}
    \centering
    \includegraphics[width=\columnwidth]{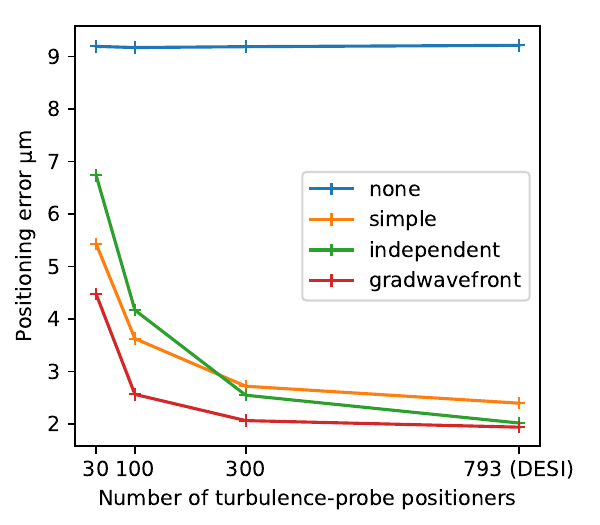}
    \caption{The positioning accuracy as a function of the number of positioners used to measure the turbulence, for a variety of different turbulence correction schemes.  Even when only 30 positioners are available to probe the turbulence, turbulence-induced errors can be reduced by half for DESI, even with a simple correction scheme.  Using a Gaussian process and taking advantage of the curl-free nature of the turbulence field provides the best performance (gradwavefront), though when a large number of turbulence probes are available (e.g., for DESI), the improvement with respect to the independent scheme is negligible.    
    }
    \label{fig:nstuckperformance}
\end{figure}

We investigate how well we can correct for the effect of turbulence as a function of the number of stationary points in Figure~\ref{fig:nstuckperformance}, for a variety of different turbulence correction schemes.

We consider four approaches for turbulence correction.
\begin{enumerate}
    \item None: no correction is performed
    \item Simple: the measured value from the nearest turbulence probe is used
    \item Independent: a Gaussian process with a squared exponential kernel is used, separately for the $x$ and $y$ components of the displacement field
    \item Gradwavefront: a Gaussian process with a squared exponential kernel is used, where a scalar field has the correlations and the observed displacements are given by its gradient.
\end{enumerate}
These are roughly intended to scale up in terms of physical accuracy and complexity.  The ``independent'' approach (3) is what is used in DESI operations.

For this test, we used the same sequence of 400 FVC images also used for \textsection\ref{subsec:correlation}.  We could therefore estimate the ``true'' position of each positioner extremely well from the mean position over the 400 exposures.  We then tested the accuracy of the different correction schemes according to how well they predicted this true position on the basis of a single FVC exposure.  We varied the number of stationary points available to the different approaches by randomly subselecting among the 793 total stationary points available to DESI to 30, 100, and 300 stationary points.

As shown in Figure~\ref{fig:nstuckperformance}, all of these approaches do well (except for the trivial ``none'' option).  With the large number of turbulence probes available to DESI, just looking up the nearest stationary point corrects positions to within 3~\mum.  The Gaussian-process based approaches do better, $\sim 2$~\mum, presumably because they are able to average over the small amount of centroid noise in the measurement of the positions of DESI fibers ($\sim 2$~\micron, \citealt{Baltay:2019}). The approach using the gradient of a wavefront has the best performance, especially when the available number of turbulence probes is small; for example, when only 100 probes are available, it achieves 2.5~\mum\ accuracy while the nearest neighbor approach gives 4~\micron.

The ``independent'' approach we actually adopted for DESI is nearly as good as the full gradient approach for the large numbers of turbulence probes available to DESI, but it actually performs worse than the simple approach when the number of turbulence probes is small.  This is because we chose to approximate the turbulence correlation function with a squared exponential kernel with a length scale of 50~mm, while a better choice would have been a kernel that falls less steeply in the wings.  Still, given the large number of turbulence probes DESI actually has, our choice performs very well, within a tenth of a micron of the best performing option.

DESI has 120 fixed fiducials intended solely for measuring the locations of different fixed points in the focal plane; this already allows the nearest-neighbor approach to reach 4~\mum\ accuracy and the gradient approach to reach 3~\mum.  Practically, this performance is more than adequate and additional turbulence probes are not needed.  In DESI we do include many more measurements, because they are available and large numbers make the system more robust.  But DESI would prioritize targeting additional astronomical sources to probing turbulence in between the primary mirror and corrector lenses.

Note that we have phrased this discussion in terms of the number of stationary points needed to probe turbulence, but ultimately what is more important is the distribution of stationary points through the focal plane.  For example, were all of the stationary points in one corner of the focal plane they would function essentially as a single measurement, and would be unhelpful for measuring the turbulence elsewhere.  Ideally the turbulence probes should be uniformly distributed across the focal plane and intermingled with the science fibers, so that they provide uncorrelated measurements of the turbulence where the science fibers reside.  For the analysis of this subsection, we have simply randomly subsampled among the DESI stationary positioners, achieving a somewhat uniform distribution but with substantial random fluctuations.

\subsection{Other Approaches for Modeling Turbulence}
\label{subsec:otherapproaches}

We expect the gradient-of-scalar-field turbulence modeling of Equation~\ref{eq:gradwavefront} to be close to optimal.  The mathematical machinery needed to compute it is very manageable, and so we recommend that approach.  However, for systems like DESI with large numbers of stationary points, \textsection\ref{subsec:howmany} makes clear that even simply looking up the nearest stationary point will give good performance.  Interpolation techniques will moreover do better than that naive approach.  Similarly, alternatives like fitting the turbulence measurements with high-order polynomials should perform well.  The gradient of a wavefront approach is most clearly required when the number of turbulence measurements is small, but when large numbers of turbulence measurements are available the problem becomes easy to solve.

Relatedly, while DESI has an oversupply of stationary points, an optimal layout conserving area in the focal plane would feature between 30 and 100 stationary points uniformly distributed through the focal plane.  As indicated in Figure~\ref{fig:nstuckperformance}, in this regime the gradient-of-wavefront approach meaningfully outperforms the simple or independent approaches we consider.

\section{Conclusions}
\label{sec:conclusion}

The Dark Energy Spectroscopic Instrument has observed millions of stars and galaxies in its first three years of operations.  These observations depend on accurately placing fibers on sources.  To accomplish this, DESI images back-illuminated fibers in the focal plane to measure their positions, and then adjusts those positions to place them closer to their intended locations.  This correction dramatically improves the accuracy of DESI's positioning, but has the potential to introduce correlated positioning errors.  These errors stem from turbulence in the volume of air through which DESI images the focal plane and the effect of that turbulence on the observed centroids of fibers.  We model and remove these turbulence-induced errors by taking advantage of stationary points in the focal plane, whose apparent motions can be attributed to turbulence.  By interpolating between and averaging these measured motions in an optimal way using a Gaussian process, we reduce the typical 7.3~\mum\ root-mean-square positioning error to about 3.5~\mum, improving the survey speed by roughly 1.6\%.

We study the correlation function of the measured centroid offsets and find that they are well modeled as the gradient of a scalar field, consistent with expectations that ultimately turbulence is distorting the wavefront of the light seen by the fiber view camera.  We find a characteristic scale of roughly 100~mm in the focal plane, which corresponds to $\approx 35$~mm halfway between the FVC and the first lens in the corrector.

We study how the performance of the turbulence correction depends on the number of stationary points available to measure the turbulence.  We find that only a relatively small number are needed to obtain good performance.  In the data set used for this study, DESI had 793 stationary points used for measuring turbulence.  Even with only 100 points, we could have corrected turbulence to $<3~\mum$, in practical terms no worse than the $\approx 2~\mum$ we obtain with the full complement.  Since DESI has 120 fiducials, we need not consider any smaller numbers, but we note that even with only 30 stationary points we could still correct turbulence to better than $<5~\mum$ with DESI-like data.  This accuracy is made possible by excellent, high signal-to-noise images from the fiber view camera and precise modeling of the centroids in those images \citep{Baltay:2019}.

Turbulence will affect other large robotically-positioned multi-object spectrographs like LAMOST, SDSS-V, PSF, and MOONS in a similar way.  We expect that these surveys would benefit from using stationary points to model and remove turbulence from the measured locations of fibers.

\begin{acknowledgements}
Chris Blake suggested using the approach of \citet{Gorski:1988} and \citet{Monin:1975} to investigate the turbulence correlation function.  His guidance was critical to \textsection\ref{subsec:correlation}.  We thank the anonymous referee for detailed feedback which significantly improved the manuscript.

SEK acknowledges support from the Science \& Technology Facilities Council (STFC)
grant ST/Y001001/1.  For the purpose of open access, the author has applied a Creative Commons
Attribution (CC BY) licence to any Author Accepted Manuscript version arising
from this submission.

This material is based upon work supported by the U.S. Department of Energy (DOE), Office of Science, Office of High-Energy Physics, under Contract No. DE–AC02–05CH11231, and by the National Energy Research Scientific Computing Center, a DOE Office of Science User Facility under the same contract. Additional support for DESI was provided by the U.S. National Science Foundation (NSF), Division of Astronomical Sciences under Contract No. AST-0950945 to the NSF’s National Optical-Infrared Astronomy Research Laboratory; the Science and Technology Facilities Council of the United Kingdom; the Gordon and Betty Moore Foundation; the Heising-Simons Foundation; the French Alternative Energies and Atomic Energy Commission (CEA); the National Council of Humanities, Science and Technology of Mexico (CONAHCYT); the Ministry of Science, Innovation and Universities of Spain (MICIU/AEI/10.13039/501100011033), and by the DESI Member Institutions: \url{https://www.desi.lbl.gov/collaborating-institutions}. Any opinions, findings, and conclusions or recommendations expressed in this material are those of the author(s) and do not necessarily reflect the views of the U. S. National Science Foundation, the U. S. Department of Energy, or any of the listed funding agencies.

The authors are honored to be permitted to conduct scientific research on Iolkam Du’ag (Kitt Peak), a mountain with particular significance to the Tohono O’odham Nation.
\end{acknowledgements}

\facilities{Mayall}
\software{astropy \citep{astropy2013a, astropy:2018, astropy:2022}}

\bibliography{turbulence}
\end{document}